\documentstyle[nolta]{article}
\input tcilatex
\QQQ{Language}{
American English
}

\begin{document}

\title{Three Body Multichannel Scattering as a Model of Irreversible Quantum
Mechanics}
\author{Alexander V. Bogdanov and Ashot S. Gevorkyan}
\date{Institute for High-performance Computing and Data Bases\\
P/O Box, 71, St-Petersburg, 194291, Russia, ashot@fn.csa.ru}
\maketitle

\begin{abstract}
The new formulation of the theory of multichannel scattering on the example
of collinear model is proposed. It is shown, that in the closed three-body
scattering system the principle of quantum determinism in general case
breaks down and we have a micro-irreversible quantum mechanics.
\end{abstract}

\section{Introduction}

All the processes, described by standard quantum mechanical approach, are
stochastic processes from the point of view of classical dynamics. The
natural equivalence between Schr\"odinger and Fokker-Plank equations was
used for formulation of quantum mechanics as stochastic theory [1], and the
procedure of quantization was introduced [2], that takes into account the
influence of stochastic processes on dynamics. For solution of quantum
problems different numerical algorithms were proposed for stochastic
dynamics (see [3]). Note, that in all above approaches the formulation of
the main quantum object, that is the wave function, was deterministic. We
must underline, that deterministic features of the physical theory are the
outcome of the symmetry of its main equations with the change of the sign of
time evolution.

At same time there is a lot of evidences for quantum deterministic
description violation both in physics (see [4] ) and in chemistry [5-6].

In present communication, based on previous research on scattering S-matrix
[7-9], it is shown, that nonstationary multichannel scattering in collinear
three-body system can be formulated as a problem of wave packet evolution in
a moving local coordinate system, that makes in general case complex, some
times chaotic, motion on the Riemann manifold (Lagrange surfaces of a
system). It is proved, that for described system the quantum determinism
breaks down and we have a typical example of irreversible quantum dynamics.

\section{Intrinsic geometry of collision system}

The Schr\"odinger equation of the collinear reactive system $A+\left(
B,C\right) _n\rightarrow $ $\left( A,B\right) _m+C$ after Delves-Smith
transformation, proposed in [10], can be reduced to

\begin{equation}
\begin{array}{c}
\left\{ \hbar ^2\Delta _{\left( x,y\right) }+2\mu _0\left[ E-V\left(
x,y\right) \right] \right\} \Psi =0 
\text{,} \\  \\ 
\Delta _{\left( x,y\right) }=\partial _x^2+\partial _y^2, 
\end{array}
\end{equation}

where $\mu _0=\left[ m_Am_Bm_C/(m_A+m_B+m_C)\right] ^{\frac 12}$ is reduced
mass of particles $m_A$, $m_B$ and $m_C$ having the total energy $E$ and $%
V\left( x,y\right) $ being the interaction potential between particles.

Let us consider the $S_p$ surface, given by parametric equation $f\left(
E_k^i;x,y\right) =-\sqrt{2\mu _0\left[ E_k^i-V\left( x,y\right) \right] }$,
where $E_k^i$ is the translational energy in $R_{in}^2$. On the $S_p$ it is
convenient to choose the curve $\Im $ (reaction coordinate) which connect
the $R_{in}^2$ and $R_{out}^2$ asymptotic sub-spaces of multichannel
scattering and that comes near to the curve $\Im _{ext}$ (the extremal rays
of Lagrange manifold of a system).

Along the curve $\Im $ let us determine the moving local coordinate system $%
\left( O\left( t\right) ,\vec e_1,\vec e_2,\vec e_3\right) _\Im $ in such a
way, that a unit vector $\stackrel{\rightarrow }{e}_1(t)$ is directed along
the tangent to $\Im $ and $\stackrel{\rightarrow }{e}_2(t)$ is orthogonal to 
$\stackrel{\rightarrow }{e}_1(t)$ and directed along the tangent of $S_p$,
oriented in unique way by determination of $\vec e_3(t)$, that can be taken
as vector product of $\vec e_1(t)$ and $\vec e_2(t)$.It should be noted,
that vectors obeying to above conditions form a manifold

\begin{equation}
M\left( \Im \right) =\stackunder{q}{\cup }\Lambda _q=\left\{ \Lambda
_q\right\} , 
\end{equation}

with $M\left( \Im \right) $ being the sum of finite or infinite number of
maps $\Lambda _q$, on each of which one can determine local coordinates $%
x_q^i$, ($i=1,2$) and coordinate transformations.

The displacement $d\stackrel{\rightarrow }{r}$ on $S_p$ is given in terms of
local coordinates $u=u\left( x,y\right) ,$ $v=v\left( x,y\right) $ and we
choose the motion to be infinite along the $u$ axis. In such a way (see [11])

\begin{equation}
\begin{array}{c}
d 
\stackrel{\rightarrow }{r}=\stackrel{\rightarrow }{e}_1\left( 1+\lambda
_1/\rho _1\right) du+\stackrel{\rightarrow }{e}_2\left( 1+v/\rho _2\right) dv%
\text{,} \\  \\ 
\lambda _1\left( u\right) =\hbar /P_0\left( u,0\right) , \\  
\\ 
P_0\left( u,v\right) = 
\sqrt{2\mu _0\left[ E-U\left( u,v\right) \right] }, \\  \\ 
U\left( u,v\right) =V\left( x\left( u,v\right) ;y\left( u,v\right) \right) , 
\end{array}
\end{equation}

and for metric tensor elements one has

\begin{equation}
\begin{array}{c}
g_{11}=\left( 1+\lambda _1/\rho _1\right) ^2 
\text{, \quad }g_{12}=g_{21}=0\text{,} \\  \\ 
g_{22}=\left( 1+v/\rho _2\right) ^2, 
\end{array}
\end{equation}

$\lambda _i$, $\rho _i$ - are the projection of de Broglie wave and main
curvature correspondingly on plane of coordinates $u$ and $v$.

If $\Im $ corresponds to $\Im _{ext}$ one has

\begin{equation}
\rho _1\left( u\right) =-p/p_u\text{,\quad }\rho _2\left( u\right) =-p/p_v, 
\end{equation}

with $p\left( u\right) =P\left( u,0\right) ,$ $p_{-}=\stackunder{%
u\rightarrow -\infty }{\lim }p\left( u\right) $ and $p_{x^i}=\partial
_{x^i}P\left( u,0\right) $, where $P\left( u,v\right) =\sqrt{2\mu _0\left[
E_k^i-U\left( u,v\right) \right] }$ and $E_k^i=p_{-}^2/2\mu _0.$

Note, that every curve $\Im $ generates its own $M\left( \Im \right) $, that
would be topologically equivalent to $M\left( \Im _{ext}\right) $,
nevertheless individual maps $\Lambda _q$ can vary significantly from one
curve to another. The choice of the curve $\Im $, different from $\Im
_{ext}, $ is supported by the fulfillment of quasiclassical conditions in
every point of it,

\begin{equation}
\begin{array}{c}
\lambda _1\left( u_q\right) \ll \left| \rho _1\left( u_q\right) \right|
,\quad 
\text{ }\lambda _2\left( u_q\right) \ll \left| \rho _2\left( u_q\right)
\right| , \\  \\ 
\text{or\quad }\lambda _1\left( u_q\right) \lambda _2\left( u_q\right) \ll
\Lambda _q\ \sim \left| \rho _1\left( u_q\right) \rho _2\left( u_q\right)
\right| . 
\end{array}
\end{equation}

where $\lambda _1\left( u_q\right) $ and $\lambda _2\left( u_q\right) $ are
the de Broglie wave lengths of imaging point by the $u$ and $v$ coordinates
correspondingly, and $\lambda _2\left( u\right) \leq v_0\left( u\right) $,
where $v_0\left( u\right) $ is the distance between the caustics.

\section{Equation of local coordinate system motion}

Let us study free motion of image point on manifold $M\left( \Im \left(
u\left( t\right) \right) \right) $.In most general case it is given by
equation (see [11])

\begin{equation}
\begin{array}{c}
x_{;tt}^k+\left\{ _{ij}^k\right\} _Mx_{;t}^ix_{;t}^j=a\left( t\right)
x_{;t}^k 
\text{,} \\  \\ 
a\left( t\right) =-t_{ss}/s_t^2,\text{ \quad }\left( i,j,k=1,2\right) 
\end{array}
\end{equation}

with $x_{;t}^k=d_tx^k,$ $x_{;tt}^k=d_t^2x^k,$ $t_{ss}=d_s^2t$ and $s_t=d_ts$.

In (7) $t$ is the natural parameter of motion, i.e. time, $s$ - is the
length of curve, $\left\{ _{ij}^k\right\} _M$ is affine connectivity on
2-D-manyfold, that is determined for nondegenerate matrix $g=\det
(g_{ij})\neq 0$ by

\begin{equation}
\begin{array}{c}
\left\{ _{ij}^k\right\} _M=\frac 12g^{kl}\left(
g_{lj;i}+g_{il;j}-g_{ij;l}\right) 
\text{,} \\  \\ 
g_{ij;k}=\partial _{x^k}g_{ij} 
\end{array}
\end{equation}

Differential equation (7) with initial conditions gives for any moment $t$
the unique solution for trajectory and speed of motion. The local coordinate
system motion is given by the equation for $x^1=u\left( t\right) $ and the
second equation for $x^2=v\left( t\right) $ is in some sense the controlling
equation. If those equations can be splitted, direct transitions are
possible in the quantum problem.

Note, that for any moment $t$ the $M\left( \Im \left( u\left( t\right)
\right) \right) $ is Riemann manyfold, but transforming the problem to
stationary representation one gets new non Riemann manyfold $M_{stc}\left(
\Im \left( u\right) \right) $, that can have curve-chaotic structure.

\section{Solution of Schr\"odinger equation on manyfold $M\left( \Im \left(
u\right) \right) $}

The Schr\"odinger equation in curvilinear coordinates can be represented as
[12]

\begin{equation}
\begin{array}{c}
\left\{ \hbar ^2g^{-\frac 12}\partial _{x^i}\left( g^{ij}\sqrt{g}\partial
_{x^j}\right) +P_0^2\left( u,v\right) \right\} \Psi =0, 
\end{array}
\end{equation}

Our purpose is to find the solution of the equation (9), that will satisfy
the following asymptotic conditions for the full wave function of the system

\begin{equation}
\begin{array}{c}
\stackunder{u\rightarrow -\infty }{\lim }\Psi ^{+}\left( u,v\right) =\Psi
_{in}\left( n;u,v\right) + \\  \\ 
\stackunder{m\neq n}{\sum }R_{mn}\Psi _{in}^{*}\left( m;u,v\right) , \\  \\ 
\stackunder{u\rightarrow +\infty }{\lim }\Psi ^{+}\left( u,v\right) =%
\stackunder{m}{\sum }S_{nm}\Psi _{out}\left( m;u,v\right) , 
\end{array}
\end{equation}

where $n$ and $m$ are the quantum numbers of the bound states in
corresponding channel.

In (10) the coefficients $R_{mn}$ and $S_{nm}$ are the excitation and
rearrangement amplitudes correspondingly.

Taking into account the fact that scattering wave function is located along
the reaction coordinate $\Im $ and based on parabolic equation method [13]
for such a problem, we represent the solution of (7) in a form

\begin{equation}
\begin{array}{c}
\Psi ^{+}\left( u,v\right) =\exp \left( i\hbar ^{-1} 
\stackrel{u}{\stackunder{0}{\int }}p(u^{^{\prime }})s_{u^{^{\prime
}}}du^{^{\prime }}\right) \times \\  \\ 
A\left( u,v\right) ,\qquad s_u=d_us. 
\end{array}
\end{equation}

After the coordinate transformation in equation (9)

$$
\tau =\left( E_k^i\right) ^{-1}\stackrel{u}{\stackunder{0}{\int }}%
p(u^{^{\prime }})s_{u^{^{\prime }}}du^{^{\prime }}\text{,\quad }z=\left(
\hbar E_k^i\right) ^{-\frac 12}pv, 
$$

taking into account (5), (6) for the total wave function of three-body
system in harmonic approximation one gets

\begin{equation}
\begin{array}{c}
\widetilde{\Psi }^{+}\left( n;z,\tau \right) =\QOVERD[ ] {\left( \Omega
_{in}/\pi \right) ^{\frac 12}}{2^nn!\left| \xi \right| }^{\frac 12}\times \\
\\ 
\exp \left[ iS_{eff}\left( z,\tau \right) \right] H_n\left[ \frac{\sqrt{%
\Omega _{in}}}{\left| \xi \right| }\cdot \left( z-\eta \right) \right] , 
\end{array}
\end{equation}

\begin{equation}
\begin{array}{c}
S_{eff}\left( z,\tau \right) =S_{cl}\left( \tau \right) -E_v^i
\stackrel{\tau }{\stackunder{-\infty }{\int }}\left( \left| \xi \right|
^{-2}-2\mu _0E_k^ip^{-2}\right) d\tau ^{^{\prime }}+ \\  \\ 
\{\dot \eta \left( z-\eta \right) +\frac 12\dot \xi \xi ^{-1}\left( z-\eta
\right) ^2-\frac 12\dot pp^{-1}z^2\}, \\  
\\ 
S_{cl}\left( \tau \right) =E_k^i\tau /\hbar -E_k^i
\stackrel{\tau }{\stackunder{-\infty }{\int }}\{\frac 12[(\dot \eta
)^2-\Omega ^2(\tau ^{^{\prime }})\eta ^2]+ \\  \\ 
F(\tau ^{^{\prime }})\eta \}d\tau ^{\prime },\qquad E_v^i=\Omega _{in}\left(
n+\frac 12\right) .
\end{array}
\end{equation}

The function $\xi \left( \tau \right) $ is the solution of classical
oscillator problem

\begin{equation}
\begin{array}{c}
\ddot \xi +\Omega ^2\left( \tau \right) \xi =0 
\text{,} \\  \\ 
\Omega ^2\left( \tau \right) =-\left( \frac{E_k^i}p\right) ^2\left[ \frac{%
p_{vv}}p+\frac{p_v^2}{p^2}+\frac 1{\rho _2^2}+\frac{p_{uu}}p+\frac{p_u^2}{p^2%
}\right] >0 
\end{array}
\end{equation}

with asymptotic condition

\begin{equation}
\begin{array}{c}
\xi \left( \tau \right) 
\stackunder{\tau \rightarrow -\infty }{\sim }\exp \left( i\Omega _{in}\tau
\right) , \\  \\ 
\xi \left( \tau \right) \stackunder{\tau \rightarrow +\infty }{\sim }c_1\exp
\left( i\Omega _{out}\tau \right) -c_2\exp \left( -i\Omega _{out}\tau
\right) , 
\end{array}
\end{equation}

with $c_1$ and $c_2$ being constants, determined by the solution of (14).

As to the function $\eta \left( \tau \right) $, it satisfies the equation

\begin{equation}
\ddot \eta +\Omega ^2\left( \tau \right) \eta =F\left( \tau \right) , 
\end{equation}

and expressed via the solution of the equation (14)

\begin{equation}
\begin{array}{c}
\eta \left( \tau \right) =\left( 2\Omega _{in}\right) ^{-\frac 12}\left[ \xi
\left( \tau \right) d^{*}\left( \tau \right) +\xi ^{*}\left( \tau \right)
d\left( \tau \right) \right] , \\  
\\ 
\eta \left( -\infty \right) =\dot \eta \left( -\infty \right) , \\  
\\ 
d\left( \tau \right) =\left( 2\Omega _{in}\right) ^{-\frac 12}\stackrel{\tau 
}{\stackunder{-\infty }{\int }}d\tau ^{^{\prime }}\xi (\tau ^{^{\prime
}})F(\tau ^{^{\prime }}), 
\end{array}
\end{equation}

Note that in the limit $\tau \rightarrow -\infty $ the exact wave function
(12) is reduced to its asymptotic form $\widetilde{\Psi }_{in}\left(
n;z,\tau \right) $

$$
\widetilde{\Psi }_{in}\left( n;z,\tau \right) =\stackunder{\tau \rightarrow
-\infty }{\lim }\widetilde{\Psi }^{+}\left( n;z,\tau \right) = 
$$

\begin{equation}
\QOVERD[ ] {\left( \Omega _{in}/\pi \right) ^{\frac 12}}{2^nn!}^{\frac
12}\exp \left( i\frac{E_k^i\tau }\hbar -\frac{\Omega _{in}}2z^2\right)
H_n\left( \sqrt{\Omega _{in}}z\right) . 
\end{equation}

Asymptotic wave function $\widetilde{\Psi }_{out}\left( m;z,\tau \right) $
is found from (16) by substitution $\Omega _{in}\rightarrow \Omega _{out}$
and $n\rightarrow m$.

\section{S-matrix representation in terms of internal time}

Let us discuss the exact representation for the S-matrix in terms of total
wave functions of (in) and (out) states [14]

\begin{equation}
\Psi ^{+}\left( n;u,v\right) =\stackunder{k}{\sum }\Psi ^{-}\left(
k;u,v\right) S_{kn,} 
\end{equation}

Taking into account, that asymptotic wave functions form a basic set for
(out) asymptotic space $R_{out}^2$, after projection of (19) onto this
asymptotic state in the limit $u\rightarrow +\infty $ one gets

\begin{equation}
\begin{array}{c}
S_{mn}= 
\stackunder{u\rightarrow +\infty }{\lim }\left\langle \Psi _{out}^{*}\left(
m;u,v\right) \Psi ^{+}\left( n;u,v\right) \right\rangle _v= \\  \\ 
=\stackunder{\tau \rightarrow +\infty }{\lim }\left\langle \widetilde{\Psi }%
_{out}^{*}\left( m;z,\tau \right) \widetilde{\Psi }^{+}\left( n;z,\tau
\right) \right\rangle _z\text{,\quad }\left\langle ...\right\rangle _z=%
\stackrel{+\infty }{\stackunder{-\infty }{\int }}dz. 
\end{array}
\end{equation}

In such a way we have a new representation for $S-$matrix, that is one
integral less, than standard one. As in this case the variable $u$ or $\tau $
plays a role of natural parameter $t$ (usually time) in scattering theory,
and (20) is quite similar to nonstationary $S-$matrix representation, we
shall follow Prigogin [15] and call $\tau $ the ''internal'' time of
three-body system.

\section{Transition amplitude for rearrangement processes}

Let us use the exact expression for $S-$matrix (20) for calculation of the
reaction transition amplitude. By using (12) for total wave function and
(17) for asymptotic one we obtain the final expression for $A+\left(
B,C\right) _n\rightleftharpoons \left( A,B\right) _m+C$ reaction amplitude

\begin{equation}
\begin{array}{c}
W_{mn}=\left| S_{mn}\right| ^2=\left( 
\frac{1-\theta }{m!n!}\right) ^{\frac 12}\left| H_{mn}\left( b_1,b_2\right)
\right| ^2\times \\  \\ 
\exp \left[ -\nu (1-\sqrt{\theta }\cos 2\o )\right] . 
\end{array}
\end{equation}

Here the function $H_{mn}\left( b_1,b_2\right) $ stands for complex
Hermitian polynomial, and

\begin{equation}
\begin{array}{c}
b_1= 
\sqrt{\nu \left( 1-\theta \right) }\exp \left( i\o \right) \text{,} \\  \\ 
b_2=- 
\sqrt{\nu }\left[ \exp \left( -i\o \right) -\sqrt{\theta }\exp \left( i\o
\right) \right] , \\  \\ 
\o =\frac 12\left( \delta _1+\delta _2\right) -\beta . 
\end{array}
\end{equation}

The variables $\theta $ , $\delta _1$, $\delta _2$, $\beta $ and $\nu $ are
determined by coefficients $c_1$, $c_2$ and by solution $\xi \left( \tau
\right) $

\begin{equation}
\begin{array}{c}
c_1=e^{i\delta _1}\left( \Omega _{in}/\Omega _{out}\right) ^{\frac 12}\left(
1-\theta \right) ^{-\frac 12} 
\text{,} \\  \\ 
c_2=e^{i\delta _2}\left( \Omega _{in}/\Omega _{out}\right) ^{\frac 12}\left[
\theta /\left( 1-\theta \right) \right] ^{\frac 12} \\  
\\ 
\theta =\left| c_2/c_1\right| ^2\text{,\quad }d=\stackunder{\tau \rightarrow
+\infty }{\lim }d\left( \tau \right) =\sqrt{\nu }\exp \left( i\beta \right)
. 
\end{array}
\end{equation}

In case of forced motion ($\nu =0$) the local coordinate system $\left(
O\left( t\right) ,\vec e_1,\vec e_2,\vec e_3\right) _{\Im _{ext}}$ is moving
along extremal path $\Im _{ext}$ and depending on one parameter $\theta $

\begin{equation}
W_{mn}=\frac{n_{\prec }!}{n_{\succ }!}\sqrt{1-\theta }\left| P_{\left(
n_{\succ }+n_{\prec }\right) /2}^{\left( n_{\succ }-n_{\prec }\right)
/2}\left( \sqrt{1-\theta }\right) \right| ^2 
\end{equation}

with $n_{\prec }=\min \left( m,n\right) $, $n_{\succ }=\max \left(
m,n\right) $ , and $P_k^n\left( x\right) $ is the associated Legendre
polynomials.

Thus, the rearrangement in collinear three-body system at overbarrier
transitions to good approximation is equivalent to the problem of quantum
harmonic oscillator with variable frequency $\Omega \left( \tau \right) $ in
the external field $F\left( \tau \right) $ with $\tau $ being the
''internal''time, that in certain conditions can be chaotic. It's in turn
means, that in general case the parameters $\theta ,$ $\delta _1,$ $\delta
_2 $ and $\nu $ and hence $S$-matrix elements being random too.

\section{Conclusion}

In present paper we tried to introduce the new formulation of the theory of
multichannel scattering on the example of collinear model. It is shown, that
there exist a natural formulation of the problem with the evolution equation
of the system determined in intrinsic geometry $M\left( \Im \left( u\right)
\right) $ on Riemann manyfold $\Im \left( u\right) $, described in term of
moving local coordinate system $\left( O\left( t\right) ,\stackrel{%
\rightarrow }{e}_1,\stackrel{\rightarrow }{e}_2,\stackrel{\rightarrow }{e}%
_3\right) _\Im $. It is proved, that motion of the local coordinate system,
obeying to the system of equations (2), on manyfold $M\left( \Im \left(
u\right) \right) $ in general case is quite complicated and sometimes
chaotic. It is shown, that transition to stationary description can be
realized by introduction of a new wave equation (9), in general case, on
curved, foam-like manyfold $M\left( \Im \left( u\right) \right) $. By means
of choosing of the corresponding reaction path coordinate $\Im $ (the curve
that connects the asymptotic scattering sub-spaces $R_{in}^2$ and $R_{out}^2$%
) along which the local coordinate system makes progressive motion, the
initial two-dimensional hyperbolic equation can be approximated at any
collision energy with any precision and reduced to one-dimensional
non-stationary equation for anharmonic quantum oscillator in a general case
with chaotic internal time $\tau \left( u\left( t\right) \right) $. In
internal time formulation the new representation of transition matrix, being
by one integration smaller, was built, and the fact that in case when
internal time $\tau $ being natural parameter $t$, it represents the exact
S-matrix of non-stationary scattering problem was proved.

In other words, the micro-irreversible quantum representation for the
problem of multichannel quantum scattering was obtained for the first time.
It means, that in the closed three-body scattering system the principle of
quantum determinism in general case breaks down.

\end{document}